\documentclass[letterpaper,twocolumn,10pt]{article}

\PassOptionsToPackage{hyphens}{url}
\usepackage{usenix2026_SOUPS}
\usepackage{xurl}

\usepackage{booktabs}         \usepackage{amsmath}          \usepackage{amssymb}
\usepackage{xfp}              \usepackage{tikz}             \usetikzlibrary{shapes.geometric, arrows.meta, positioning}
\usepackage{graphicx}         \usepackage{subcaption}

\newcommand{\sm}[1]{\textcolor{blue}{}}

\renewcommand{\paragraph}[1]{\vspace{-1em}{\hfill \break \textbf{#1.}}}

\newif\ifanonymous
\anonymousfalse

\title{\Large \bf A Penny for Your Prompts: \\ Experiments Detecting and Mitigating LLM Usage by Survey Respondents}
\def\plainauthor{Zane Xu and Nathan Malkin}  

  \author{
    {\rm Zane Xu}\\
    New Jersey Institute of Technology \\
    \and
    {\rm Nathan Malkin}\\
    New Jersey Institute of Technology \\
  }

\date{}

\begin{document}
\maketitle
\thecopyright

\newcommand{\percentageMBLLM}{27.5}
\newcommand{\percentageMBSelf}{31.4}
\newcommand{\percentageMbots}{40.8}
\newcommand{\MBnumber}{51}
\newcommand{\MBdisablenumber}{7}
\newcommand{\percentagedisable}{\fpeval{round(\MBdisablenumber/\MBnumber*100,1)}\unskip
}
\newcommand{\Mnumber}{49}
\newcommand{\MLLMnumber}{41}
\newcommand{\percentageMLLM}{\fpeval{round(\MLLMnumber/\Mnumber*100,1)}\unskip
}
\newcommand{\MBlownumber}{15}
\newcommand{\percentagelow}{\fpeval{round(\MBlownumber/\MBnumber*100,1)}\unskip
}
\newcommand{\Pnumber}{50}
\newcommand{\PLLMnumber}{4}
\newcommand{\tenPLLMnumber}{1}
\newcommand{\kindPLLMnumber}{0}
\newcommand{\percentagePllm}{\fpeval{round(\PLLMnumber/\Pnumber*100,1)}\unskip
}
\newcommand{\percentagetenPllm}{\fpeval{round(\tenPLLMnumber/\Pnumber*100,1)}\unskip
}
\newcommand{\percentagekindPllm}{\fpeval{round(\kindPLLMnumber/\Pnumber*100,1)}\unskip
}
\newcommand{\Alan}{6}
\newcommand{\Leonardo}{5}
\newcommand{\Einstein}{2}
\newcommand{\Allllmnumber}{46}
\newcommand{\percentagellmcorrect}{67.4}
\newcommand{\percentagenonllmcorrect}{90.9}

\newcommand{\surveyandparticipantstable}{\begin{table}[t]
		\centering
		\caption{Conditions and participant distribution}
		\label{tab:five_surveys}
		\small  \begin{tabular}{lrcr}
			\toprule
			Condition          & Time (min) & Platform & $N$ \\
			\midrule
			Baseline           & 5          & Prolific & 50  \\
			Baseline           & 5          & MTurk    & 49  \\
			Longer version     & 10         & Prolific & 50  \\
			Asked to avoid AI  & 5          & Prolific & 50  \\
			Disable copy/paste & 5          & MTurk    & 51  \\
			\bottomrule
		\end{tabular}
	\end{table}}

\newcommand{\participantsnumber}{199}
\newcommand{\percentagellmred}{41.3}
\newcommand{\percentagenonllmred}{5.9}
\newcommand{\mediandinnerllm}{four}
\newcommand{\mediandinnernonllm}{two}

\newcommand{\percentagenonllmactor}{19.0}
\newcommand{\percentagenonllmpersonal}{39.9}
\newcommand{\llmactorN}{one}

\newcommand{\llmscientistnumber}{16}
\newcommand{\nonllmscientistnumber}{1}
\newcommand{\percentagellmpersonal}{28.3}
\newcommand{\llmopencodenumber}{4}
\newcommand{\nonllmopencodenumber}{1}

\newcommand{\pilotnumber}{88}

\newcommand{\krippendorff}{.718}
\newcommand{\percentagehumanobservedLLM}{17.9}
\newcommand{\humanjudgmentprecision}{80.0}
\newcommand{\humanfn}{11}
\newcommand{\humanfp}{6}
\newcommand{\kupper}{.65}

\newcommand{\surveyclosewordformat}{nine}
\newcommand{\surveyclosesignificant}{six}

\newcommand{\LLMembedmean}{0.535}
\newcommand{\nonLLMembedmean}{0.347}
\newcommand{\LLMJaccard}{.125}
\newcommand{\nonLLMJaccard}{.099}
\newcommand{\LLMdelete}{11.1}
\newcommand{\nonLLMdelete}{29.6}

\newcommand{\medianlengthllm}{31}
\newcommand{\medianlengthnonllm}{12}

\newcommand{\timellmmediantotal}{606.5}
\newcommand{\timenonllmmediantotal}{372.0}

\newcommand{\mousellm}{71.8}
\newcommand{\mousenonllm}{69.1}

\newcommand{\tabaccuracy}{79.9}
\newcommand{\tabfone}{.545}

\newcommand{\Prolificparticipants}{150}
\newcommand{\overallnumber}{250}
\newcommand{\MTurkallparticipants}{100}

\newcommand{\Redllmsphere}{80.4}
\newcommand{\Rednonllmround}{68.6}

\newcommand{\geminieachnumber}{10}
\newcommand{\claudeeachnumber}{10}
\newcommand{\claudekindnumber}{30}
\newcommand{\cometeachnumber}{10}
\newcommand{\claudeoptional}{six}
\newcommand{\perplexitykindnumberA}{seven}
\newcommand{\perplexitykindnumber}{30}
\newcommand{\trailall}{90}
\newcommand{\claudeopinionA}{nine}
\newcommand{\claudeevasiveA}{three}
\newcommand{\cometevasiveA}{two}
\newcommand{\claudeopinion}{ten}
\newcommand{\cometcompleteall}{28}
\newcommand{\cometselfreport}{27}
\newcommand{\cometreportGPT}{one}
\newcommand{\claudecompleteall}{35}
\newcommand{\claudeselfreport}{33}
\newcommand{\claudereportGPT}{once}
\newcommand{\cometoptional}{once}
\newcommand{\claudeeachnumberanswerB}{five}

\newcommand{\DemographicsTableCombined}{\begin{table}[h]
		\caption{Demographics of Participants}
		\label{tab:demographics}
		\centering
		\small
		\begin{tabular}{@{}lrr@{}}
			\toprule
			                                & \textbf{Prolific} & \textbf{MTurk} \\
			\midrule
			\textbf{Gender}                 &                   &                \\
			\quad Woman                     & 49\%              & 52\%           \\
			\quad Man                       & 49\%              & 48\%           \\
			\quad Non-binary/other          & 2\%               & 0\%            \\
			\addlinespace
			\textbf{Age}                    &                   &                \\
			\quad Mean (Median)             & 44 (44)           & 36 (35)        \\
			\quad Range                     & [20, 81]          & [25, 64]       \\
			\addlinespace
			\textbf{Race/Ethnicity}         &                   &                \\
			\quad White                     & 77\%              & 95\%           \\
			\quad Black or African American & 11\%              & 2\%            \\
			\quad Asian                     & 7\%               & 2\%            \\
			\quad Hispanic or Latino        & 3\%               & 0\%            \\
			\addlinespace
			\textbf{Community}              &                   &                \\
			\quad Suburb                    & 44\%              & 28\%           \\
			\quad Large city                & 27\%              & 51\%           \\
			\quad Small town                & 19\%              & 9\%            \\
			\quad Rural                     & 9\%               & 12\%           \\
			\addlinespace
			\textbf{Technology Advice}      &                   &                \\
			\quad Very frequently           & 8\%               & 29\%           \\
			\quad Frequently                & 17\%              & 32\%           \\
			\quad Sometimes                 & 43\%              & 32\%           \\
			\quad Rarely                    & 25\%              & 6\%            \\
			\quad Never                     & 7\%               & 1\%            \\
			\bottomrule
		\end{tabular}
	\end{table}
}

\newcommand{\onecp}{30.7}
\newcommand{\percentagemturkselfreport}{38.8}
\newcommand{\mturkflagpercentage}{89.5}

\newcommand{\RecommendTableAgent}{
	\begin{table}[t]
		\centering
		\scriptsize
		\caption{Number of successful survey completions by agents}
		\label{tab:agent_recommendations}
		\begin{tabular}{@{}p{4cm}rrr@{}}
			\toprule
			\textbf{Condition}                   & \textbf{Claude} & \textbf{Gemini} & \textbf{Perplexity} \\ \midrule
			Baseline (minimal)                   & 5 / 10          & 0 / 10          & 10 / 10             \\
			Baseline (opinionated)               & 9 / 10          & 0 / 10          & 9 / 10              \\
			Baseline (evasive)                   & 3 / 10          & 0 / 10          & 2 / 10              \\ \midrule
			Instructional guidance (minimal)     & 0 / 10          & 0 / 10          & 3 / 10              \\
			Instructional guidance (opinionated) & 0 / 10          & 0 / 10          & 3 / 10              \\
			Instructional guidance (evasive)     & 0 / 10          & 0 / 10          & 1 / 10              \\ \midrule
			Prohibit copy-pasting (minimal)      & 6 / 10          & 0 / 10          & 0 / 10              \\
			Prohibit copy-pasting (opinionated)  & 9 / 10          & 0 / 10          & 0 / 10              \\
			Prohibit copy-pasting  (evasive)     & 3 / 10          & 0 / 10          & 0 / 10              \\
			\bottomrule
		\end{tabular}
		\vspace{2pt}
		\raggedright
		\footnotesize
	\end{table}
}

\newcommand{\DinnerTable}{\begin{table}[h]
		\centering
		\caption{Distribution differences between suspected LLM and human responses to the dinner question}
		\label{tab:dinner_llm_nonllm_percent}
		\begingroup
		\setlength{\tabcolsep}{3pt}
		\renewcommand{\arraystretch}{1.0}
		\small
		\begin{tabular}{lrr}
			\toprule
			\ Category          & LLM    & Human  \\
			\midrule
			Science-related     & 34.8\% & 0.7\%  \\
			Personal connection & 28.3\% & 39.9\% \\
			Politician          & 13.0\% & 9.2\%  \\
			Entrepreneur        & 6.5\%  & 5.9\%  \\
			Writer              & 2.2\%  & 3.3\%  \\
			Performer           & 2.2\%  & 19.0\% \\
			Athlete             & 0\%    & 5.2\%  \\
			Historical figure   & 0\%    & 6.5\%  \\
			Influencer          & 0\%    & 1.3\%  \\
			Religion            & 0\%    & 4.6\%  \\
			Others              & 13.0\% & 4.6\%  \\
			\bottomrule
		\end{tabular}
		\endgroup
	\end{table}}
 \begin{abstract}
	Large language models are increasingly used by participants on crowdsourcing platforms when responding to surveys,
	potentially undermining the validity of collected data.
	Our study aims to quantify the prevalence of this behavior and investigate methods to detect and prevent it.
	In a series of surveys ($N = \overallnumber$), we examined conditions such as
	platform choice, survey length, requests not to use AI, and disabling copy-paste functionality.
	We were able to identify distinct characteristics of LLM-assisted responses
	and found that their frequency varied widely, from under 10\% on Prolific to over 80\% on Mechanical Turk.
	Mitigation measures reduced LLM usage but did not necessarily improve data quality.
	No participants employed browser-use agents at the time of our survey, but we report on our own detection experiments.
	We recommend that researchers actively screen survey responses for LLM usage by recording and analyzing keystroke data and crafting instructions and questions aimed at AI.
\end{abstract}

\section{Introduction}
Online surveys are an important instrument in scientific research,
and for human-centered security in particular,
but it is unclear to what extent they can remain trusted and useful going forward.
The culprit is the widespread adoption of large language models (LLMs), which are highly adept at text generation and question answering,
making responding to survey questions a relatively simple task for them.
Many anecdotal reports and recent research~\cite{Zhang25b,Veselovsky25,Zhang25c} suggest that study participants are increasingly using LLMs for this purpose.

Respondents may turn to LLMs for a variety of reasons,
ranging from improving their own writing to wholesale answer generation.
Unfortunately, those on the receiving end have no way to tell apart these use cases.
While editing assistance would not impact a study's results,
answers written by an LLM without human input are a serious issue for research validity.
Survey-based research aims to capture the opinions of real people in the target population, not the confabulations of a language model.
In addition, LLMs' ``opinions'' have been shown to exhibit a lack of diversity and systematic biases~\cite{Santurkar23,Tjuatja24a,Dominguez-Olmedo24,Doshi24,Zhang25b}.
This presents a particular problem for the human-centered security field, where
surveys can be found in a large fraction of publications at venues like SOUPS
and are used for
determining privacy preferences and contextual norms, understanding the prevalence of and reasons for security behaviors, and driving interface design decisions.

Respondents who fill out political, institutional, or product surveys may see their opinions effect change,
but participants in scientific surveys accrue no direct benefits.
With compensation as their only reward, some may be content to complete a survey as quickly as possible, without regard for the content of their answers.
This is especially common on crowdsourcing platforms like Amazon Mechanical Turk, where
data quality issues have long predated AI tools~\cite{Peer22,Chandler14,Keith17}.

Existing data quality measures are largely not up to the task.
LLMs can answer typical attention checks
(``select disagree for the third option below'')
and comprehension check questions,
arguably better than humans can~\cite{Lebrun24,Westwood25}.
LLM detectors have been repeatedly shown to be unreliable even for longer texts~\cite{Liang23a,Tang24},
and survey responses are usually relatively short,
making them even more difficult to detect~\cite{Wu24}.

Recent publications have started to raise the alarm about the risks to survey studies from LLMs~\cite{Cox24,Lebrun24,Westwood25,Schirra25,Traylor25,Zhang25b,Zhang26,Caven25}.
There have been some early attempts to quantify LLM usage~\cite{Veselovsky23} and suggestions for ways to filter or prevent it~\cite{Veselovsky25,Caven25}.
However, existing guidance is not entirely reliable (e.g., machine learning classifiers~\cite{Lebrun24}) or effective (traditional attention check questions~\cite{Westwood25}).
It may face even further challenges with the advent of LLM-based browser automation (agents).

To address these gaps, we formulated the following research questions to guide our study:

\begin{itemize}
	\setlength{\itemsep}{0pt}
	\item \textbf{RQ1: How can LLM usage in surveys be detected?} We compared the effectiveness of behavioral patterns (e.g., keystroke data), human perception, attention check questions, and self-report questions.

	\item \textbf{RQ2: How does LLM usage affect survey response characteristics?} We examined differences between LLM-generated and human responses in comprehension accuracy, open-ended themes, and privacy preferences.

	\item \textbf{RQ3: What factors influence participants' likelihood of using LLMs?} We tested the impact of crowdsourcing platforms, survey length, requests not to use AI, and restrictions on copying and pasting.

	\item \textbf{RQ4: How do browser-use agents interact with online surveys?}
	      We investigated whether the detection and mitigation strategies from RQ1--RQ3 remain applicable.
\end{itemize}

To answer these research questions, we conducted a survey-based study testing five different conditions ($N = \overallnumber$) on two major crowdsourcing platforms, Prolific~\cite{ProlificEasilycollecthighqualitydatarealpeople} and Amazon Mechanical Turk~\cite{AmazonMechanicalTurka}.
Unlike previous studies, which used a summarization task to detect LLM usage~\cite{Veselovsky25}, or directly asked participants about their LLM usage patterns~\cite{Zhang25b}, we designed a realistic survey representative of actual human-centered privacy research.
It focused on privacy preferences for an emerging technology---drones---and contained common elements like
multiple-choice and open-ended questions, comprehension checks, and attention check questions.

To evaluate detection methods, we collected metadata including keystrokes, tab-switching, mouse activity, and timing, along with survey responses.
Our results show that browser events---in particular, copying and pasting---provide a relatively reliable way of detecting LLM usage.
Suspected LLM answers are longer, more detailed,
and exhibit distinctive response patterns on simple attention check questions.
Platform choice matters:
participants on MTurk are significantly more likely to use LLMs. In contrast, survey length and explicit instructions against LLM usage had less impact.
Interestingly, restricting copy-paste resulted in lower data quality on MTurk.

Our data collection predated the general availability of browser-use agents,
so we conducted our own experiments with Claude, Gemini, and Perplexity.
The agents' behavior varied, but they were often honest in self-identifying as AI
and refusing to answer surveys that were not intended for them.
Based on these findings, we propose practical recommendations to mitigate the effect of LLMs in online surveys, such as providing explicit instructions and questions about AI usage, implementing keystroke tracking for detection, and including open-ended attention check questions.

\section{Related Work}
We organize our related work based on research questions:
LLM detection, response characteristics, and mitigation.

\subsection{Detecting and quantifying LLM usage}

Detecting LLMs in survey contexts is challenging.
Multiple studies have pointed out that LLMs can bypass traditional survey quality controls, such as attention check questions and CAPTCHAs~\cite{Westwood25, Shahania25, Hohne25}.
Fiedler and D\"{o}pke found that human raters only slightly outperform chance when detecting LLM-generated texts~\cite{Fiedler25}.
Similarly, a study by Radivojevic et al.\ found that people were not good at distinguishing LLM-generated texts in the social media context~\cite{Radivojevic24}.

Approaches using machine learning have also not succeeded.
For example,
Lebrun et al.\ claimed the automatic AI detection systems were completely unusable~\cite{Lebrun24},
while
Wu et al.\ showed that
even the state-of-the-art detectors perform unreliably in realistic settings~\cite{Wu24}.
In contrast, Claassen et al.\ were more optimistic, but their approach required longer, narrative responses and per-question model fine-tuning~\cite{Claassen26}.

Nonetheless, researchers have sought to quantify the frequency of LLM usage in surveys.
Zhang et al.\ reported that 34\% of participants self-reported LLM use in an online survey, but did not combine this with behavioral signals to verify accuracy~\cite{Zhang25b}.
Veselovsky et al.\ estimated that 33--46\% of MTurk workers used LLMs in a text summarization task using a classifier and keystroke logging, but their task was not representative of typical survey research~\cite{Veselovsky23,Veselovsky25}.
Stafford et al.\ identified roughly 21\% of focus group participants as potentially using LLMs based on text similarity~\cite{Stafford24}.
Using a similar approach, Zhang et al.\ estimated that roughly 30\% of open-ended responses in post-ChatGPT crowdsourced surveys were AI-generated~\cite{Zhang25c}.
Rilla et al.\ observed up to 45\% of participants copying or pasting text, which they suggested could be indicative of LLM use, though these are upper-bound indicators that cannot confirm actual LLM involvement~\cite{Rilla25}.

Finally, some studies have covered LLM usage without quantifying it.
Meem et al.\ reported widespread AI-generated responses in a software engineering survey~\cite{Meem24}, and
Schirra et al.\ interviewed 17 online study participants and found they used LLMs to polish rather than create responses, but did not estimate how common this was in the broader population~\cite{Schirra25}.

\subsection{LLM-generated response characteristics}
LLMs do not exhibit human-like behavior in survey contexts~\cite{Tjuatja24a} and show substantial misalignment with diverse demographic groups' opinions~\cite{Santurkar23}.
Their answers are less diverse and more positive~\cite{Zhang25b},
and are more aligned with certain political views~\cite{Hartmann23,Rozado24a}.

Beyond content differences, LLM-generated texts also differ structurally from human-written ones.
Writing by LLMs is more verbose~\cite{Kabir24}, more likely to have better logical structure, and has less radical emotion expression~\cite{Munoz-Ortiz24}.
They are also more information-dense~\cite{Reinhart25}.
Additionally, LLM writing leans toward Western style and diminished cultural nuances~\cite{Agarwal25}.
Interestingly, LLMs may influence human writing:
Yakura et al.\ observed that people have become more likely to use words commonly used by LLMs (e.g., delve)~\cite{Yakura25}, which may further complicate detection efforts.

\subsection{Factors influencing LLM usage}
Veselovsky et al.\ investigated several methods to prevent LLM usage, including banning copy-and-paste and turning question text into images~\cite{Veselovsky25}.
They found that these measures reduced estimated LLM usage but did not fully eliminate it.
Prior to the advent of LLMs, Jacquemet et al.\ pointed out oath-taking could reduce the likelihood of lying during an experiment~\cite{Jacquemet19},
which might also be effective for LLM usage.

Choice of crowdsourcing platform may also play a role.
Currently, Mechanical Turk (MTurk) faces serious concerns over LLM-generated responses~\cite{Traylor25},
but its data quality challenges extended beyond LLM usage alone,
potentially making LLM detection even more difficult.
Peer et al.\ found that MTurk had more serious data quality issues than other crowdsourcing platforms (e.g., Prolific, CloudResearch)~\cite{Peer22}, with multiple studies confirming that the data quality on MTurk was concerning~\cite{Kay25, Webb24, Shimoni25,Tang22,Ahler25}.

\subsection{Research gap}
Several research gaps remain.
First, existing detection methods have primarily evaluated LLM-generated text produced by researchers~\cite{Fiedler25, Lebrun24}, rather than responses from actual LLM users in online surveys.
Previous studies also have not combined behavioral data (e.g., tab-switching) with response data (e.g., attention check questions).
Second, prior work documented LLM response biases using researcher-generated data~\cite{Rozado24a, Zhang25b}, but these biases have not been examined with actual survey participants.
Third, while previous studies documented low data quality on MTurk~\cite{Peer22, Shimoni25}, they did not directly compare LLM usage rates across crowdsourcing platforms.
Finally, prevention strategies such as oath-taking~\cite{Jacquemet19} were studied before the LLM era, and copy-paste restrictions~\cite{Veselovsky25} were evaluated only in a summarization task, which may not generalize to typical crowdsourcing research.
Our work addresses these gaps by combining multiple detection heuristics in a human-centered security and privacy survey across Prolific and MTurk, testing mitigation strategies, and extending the analysis to browser-use agents that existing methods were not designed to handle.

\section{Methods}
The goal of our study was to detect and understand how survey respondents use LLMs.
For the purposes of this study, we define LLM usage as at least one open-ended answer in the survey being written with the assistance of an LLM.
(Participants may use LLMs in other ways, such as to better understand the survey questions, but we consider this out of scope.)
Our definition does not distinguish how the LLMs are used---for generating whole responses or cleaning up ideas and grammar.
This is because the central problem for researchers is that these two scenarios are indistinguishable.

\subsection{Study design}

The basic mechanics of our study were to conduct several surveys, collect as much data as possible, and analyze this data to check for LLM usage.

\paragraph{Survey choice}
A major design question for us was which survey to use as the basis of our experiments.
Our main criterion was that the survey should be in human-centered security or privacy,
since we wanted to study our own research context.
We initially explored replicating a prior survey from literature,
but struggled to identify a suitable choice.
Some survey papers were missing a complete survey instrument
or had recruited from narrow populations.
Many had few open-ended questions, lacked item-level results, or reported the answers qualitatively, precluding systematic statistical comparison.

Ultimately, we decided to create our own survey,
designing it to be representative of the topics and questions that may be found in typical human-centered privacy research.
We chose an emerging technology---drones---and asked participants how they felt about their data being potentially collected and what safeguards they thought this data should have.

\paragraph{Survey details}
We designed our survey to be completed in five minutes,
reasoning that much shorter surveys may not motivate participants to use an LLM,
while much longer ones would cost more without yielding majorly different insights.
However, we did hypothesize that longer surveys may induce more LLM usage,
so we designed a separate study condition that tested a longer---10-minute---survey.

The survey (and its extended variant) consisted of a mix of open-ended and multiple-choice questions, as well as some scattered expository paragraphs.
It incorporated two attention check questions and a comprehension check.
Participants were asked a standard set of demographic questions and to self-report whether they had used an LLM as part of their survey.
The complete survey instrument is in Appendix~\ref{detecting-llms-survey}.

\surveyandparticipantstable

\paragraph{Conditions}
In addition to the two length conditions, our study tested three other hypotheses. To investigate interventions that may discourage LLM usage,
we tested a variant of the 5-minute survey
in which participants were asked not to use LLMs (but this was not enforced),
and another in which copying and pasting was disabled on survey pages.
Additionally, to examine the effects of the participant recruitment platform,
we ran different surveys on Prolific~\cite{ProlificEasilycollecthighqualitydatarealpeople} and Amazon Mechanical Turk~\cite{AmazonMechanicalTurka}.
Table~\ref{tab:five_surveys} summarizes all five conditions and their participant counts.
\subsection{Data collection and analysis}
\label{meth:ana}

We hosted our survey on our university's servers using an instance of the commonly-used LimeSurvey software~\cite{LimeSurveyFreeOnlineSurveyTool}.
In addition to responses, we used JavaScript to collect data about timing, tab-switching events, mouse activity, and keystrokes, including copy and paste events and the content copied or pasted.
Additionally, we collected IP addresses to detect potential fraudulent activity from duplicate accounts.

To determine which responses were associated with LLM usage, we manually inspected all data and metadata in order to create the heuristics
we discuss in \S\ref{sec:results-rq1}.
When comparing the metrics across conditions, we use statistical tests
including Mann-Whitney~$U$~\cite{Mann47}, Chi-square~\cite{Pearson92}, and Fisher's exact test~\cite{Fisher35}.
To mitigate the risk of false positives from multiple testing, we applied the \v{S}id\'ak correction~\cite{Sidak67} to control the family-wise error rate across all comparisons.
All reported $p$-values in this paper are \v{S}id\'ak-corrected.

One of the open-ended questions in the survey was analyzed by human raters using qualitative coding~\cite{Saldana25}.
Two researchers read through responses and then constructed independent codebooks.
After that, they discussed and combined the codebooks.
Researchers then independently recoded all responses.
Because multiple codes per response were allowed,
we computed inter-rater reliability using the Kupper-Hafner concordance coefficient~\cite{Kupper89}, obtaining $\kupper{}$.
After that, two researchers resolved conflicts, achieving 100\% final agreement.

Separately, two researchers independently rated each response on three 5-point Likert scales: level of detail, language polish, and overall likelihood of LLM usage.
We calculated inter-rater agreement using Krippendorff's~$\alpha$~\cite{Krippendorff18},
though in this case the statistics represent findings about whether two humans can agree on whether a response is LLM-generated (\S\ref{sec:results-openended}).
We averaged the two raters' scores for each response and interpreted a rating above 3 as suspicion of the response.

\subsection{Recruitment}
We recruited participants between July and August 2025 on Prolific~\cite{ProlificEasilycollecthighqualitydatarealpeople} and Mechanical Turk~\cite{AmazonMechanicalTurka}.
All participants were over 18 years old, lived in the U.S.,
and had an approval rate over 98\% on their platform.\footnote{MTurk participants are required to have at least 100 approved HITs to ensure their approval rate is accurate on that platform.}
We aimed for 50 participants per study condition,
because anecdotal evidence and prior research estimated LLM usage between 5\% and 35\%~\cite{Zhang25b,Veselovsky25, Zhang25c},
and this sample size would ensure sufficient instances to run statistical tests.
Our final participant count was \overallnumber{};
Table~\ref{tab:five_surveys} shows their distribution across five conditions.

\subsection{Ethics statement}
Our consent form informed participants that they were taking part in a study about understanding people's technology-related attitudes and survey participation.
We explicitly told participants that we would record their IP addresses and their interaction with the webpage, including what they typed, where they clicked, and mouse movements.
We compensated everyone
regardless of LLM suspicion,
\$1.25 for the 5-minute survey and \$2.50 for the 10-minute survey.
Our IRB reviewed and approved all study procedures.

\subsection{Pilot study}
We conducted a pilot study
from April to May 2025
with \pilotnumber{} student volunteers who were randomly assigned to (not) use an LLM.
Our intent was to collect ground truth data to help train potential LLM detectors,
but we found that the controlled in-person environment is different from real online crowdsourcing platforms.
Some behaviors that occur frequently among online participants (e.g., browsing other websites during surveys) were not observed in the pilot.
Additionally, some participants did not shorten the long (multi-paragraph) LLM responses, making them easily stand out.

Despite these limitations, we observed different answering patterns for the attention check question between LLM users and non-users.
This finding inspired us to add a second attention check.
(Results from both are discussed in \S\ref{sec:results-attention-check}.)

\subsection{Limitations}
\label{sec:limitations}

We acknowledge a number of limitations in our study.
One issue is the narrow scope of the survey at the center of our study:
it is about a single topic of people's privacy perceptions,
whereas survey fraud affects scientific surveys of all types.
However, as usable security researchers, we wanted to know how LLM usage affected studies in our field,
and the narrower focus allowed us to craft a survey we believe to be broadly representative.
We do not claim our study generalizes to other fields,
but our results did not lead us to believe that the survey subject affected participants' choices to use LLMs
and their reactions to the interventions.

Our study had five conditions,
including survey length, recruitment platform, and intervention strategies,
but not all treatments were replicated across all conditions.
For example, we did not test disabling copy-paste features on Prolific,
because the ``softer'' version that requested participants avoid AI already resulted in no LLM usage (\S\ref{sec:results-kindlyasking}).
Conversely, we did not test the longer version of our survey on MTurk,
because we observed a very high incidence of LLM usage there even in the baseline condition
(\S\ref{sec:results-platforms}).
Similarly, because we felt the platform's data quality effects would dominate,
we chose not to test (unenforced) requests not to use AI on MTurk.
While this study design precludes us from directly comparing these specific interventions across platforms,
we are still able to compare the two platforms through the baseline conditions.

Trying to measure and understand LLMs is challenging due to the rapid pace at which this technology is evolving.
Between the time we collected the study data and when we finished analyzing it,
all major LLM providers released new models.
Each release changes the model's outputs and may in turn affect how people use them.
Nonetheless, we believe that our study yields novel insights not just about the models but about how people use them,
which is a longer-lasting finding.
New AI modalities may emerge and be adopted;
for example, browser-use agents were not widely available at the time of our data collection,
prompting us to study them separately in~\S\ref{agent}.
However, the chatbot interface has lasted several years and is widely popular,
so it is unlikely to be deprecated soon.
In the meantime, researchers need to conduct surveys now.

As with many survey studies, ours would benefit from a larger sample size,
though ours was sufficient to create statistically significant findings.
We recruited exclusively U.S. participants, as many privacy studies do to ensure a culturally homogeneous sample,
and because Prolific and MTurk are U.S.-centered platforms,
but a more global participant pool might exhibit different LLM usage,
as could other crowdsourcing platforms
(e.g., CloudResearch~\cite{CloudResearchOnlineResearchPlatform}).

Finally,
we lacked a ground truth for determining whether any responses were truly LLM generated.
Our detection heuristics are therefore not definitive
(each has its own limitations, discussed separately in \S\ref{sec:results-rq1})
but we discuss how they might still be useful in \S\ref{sec:discussion}.
\subsection{Participants and demographics}
We recruited {$\Prolificparticipants$} participants from Prolific across three survey conditions, and {$\MTurkallparticipants$} participants from MTurk across two survey conditions (see Table~\ref{tab:five_surveys}).
Both platforms showed gender balance and were predominantly white.
Compared to Prolific participants, MTurk participants were younger and more willing to provide technology advice.
Full demographic breakdowns are provided in Appendix~\ref{app:demographics} (Table~\ref{tab:demographics}).

\section{Results}
\label{sec:results}

This section reports on our experiments to detect when respondents use LLMs (\S\ref{sec:results-rq1}), understand characteristics of these responses (\S\ref{sec:results-rq2}), and affect their frequency (\S\ref{sec:results-rq3}).

\subsection{RQ1: How can LLM usage be detected?}
\label{sec:results-rq1}

After manually reviewing each response and all associated metadata,
we derived the following detection heuristics.

\subsubsection{Copy \& paste behavior}
\label{sec:results-copypaste}

Prior work~\cite{Rilla25} and our pilot study identified a key marker of LLM use: copy-pasting.
This makes sense, as it is the easiest way to give input to an LLM and provide its response.
A single copy or paste event is not necessarily indicative
(\onecp{}\% of all\footnote{
	Except where explicitly noted, all analyses
	are done with $N=\participantsnumber{}$
	and exclude the condition where copy-pasting was disabled,
	discussed in \S\ref{sec:results-bancopypaste},
	due to the significantly altered behavior that condition induced.
}
responses had at least one of these).
However, repeated copying and pasting is suspicious.

We therefore derived our first heuristic for LLM-assisted responses:
participants who
\textbf{copy survey text at least once \emph{and} paste external text at least once}.
We additionally required that the text was at least 10 words long, as shorter instances are more likely to be minor edits rather than complete responses.
By focusing on copied survey text, we exclude participants who may copy their own written responses, for example to run them through an external grammar checker.
The focus on pasted external text excludes participants who may copy phrases or sentences from the survey itself into their response.

Modern LLMs are multimodal and accept input not only as text but as images~\cite{ImageunderstandingGeminiAPI,26b}.
A user can take a screenshot of a survey page and ask the LLM to generate a response.
This would not be caught by the above heuristic, since no copying is involved.
However, they would still need to insert the LLM-generated response into the survey page.
This motivated us to develop our second heuristic:
participants who
\textbf{paste at least once per open-ended response question}.
(Researchers may choose a lower bar, such as $\tfrac{2}{3}$ of responses, but manual inspection suggested the higher one still worked in our study.)

The exact text that is being pasted can also serve as a clue to the origin of the content.
In one instance, we observed that the pasted text included ``Ask ChatGPT,'' a message that appears as a pop-up when a user selects a portion of ChatGPT's response in the interface.
(Notably, the participant removed these words before submitting the response.)
Thus,
\textbf{LLM-related strings in pasted text}
can also be used as a detection heuristic.
For example, Grok also has a pop-up when a response is selected, but it says ``Add to chat''~\cite{Grok}.

In our baseline condition, the 5-minute survey published on the Prolific platform,
{\percentagePllm}\% of participants
were flagged by a combination of the above heuristics.

The copy-paste heuristics we derived may be subject to misclassifying real human responses as LLM responses.
Some people may select and copy text without the intention of feeding it to an LLM, for example to look up unfamiliar terms.
However, they would not be flagged unless they also pasted something external.
Respondents may prefer to compose all their responses in an external editor before pasting them in.
We believe this is uncommon, as browser fields have conveniences like spell check built in.
As a mitigation, survey creators can explicitly instruct participants against doing this.

Conversely, we might also misclassify LLM-generated responses as authentic human responses.
Our heuristic would not detect a participant who provides image input to an LLM and uses its output for only a small number of questions.
We would also miss anyone who manually retyped an LLM-generated response into the form field,
though this would not save time in comparison to filling out the survey themselves,
so it is unclear why someone would do that.

Respondents may use browser extensions and integrations rather than pasting LLM responses themselves.
Browser-use agents are one such tool, which we explore separately in \S\ref{agent}.
More adversarial designs are possible, which would seek to fully mimic human typing and evade detection.
Our analysis of keystroke timings did not identify anything suspicious,
and our web searches did not find relevant tools or discussion of them.
Therefore, we do not believe they are presently common
but acknowledge that this may change in the future.
\subsubsection{Self-reported LLM usage}
\label{sec:results-selfreport}

At the end of our survey, we asked participants to
\textbf{self-report if they had used an LLM} with the question
``\textit{Did you use an LLM or other AI to help you fill out this survey? Note: your answer will not affect your compensation for the survey}.''
In our three Prolific conditions, all participants claimed not to have used an LLM.
In contrast, in the MTurk conditions, \percentagemturkselfreport{}\% of participants self-reported using an LLM.
Of these, \mturkflagpercentage{}\% had already been flagged by the copy-paste heuristic.

It is possible for participants to claim that they used an LLM without having actually done that.
However, doing so is itself a signal of poor data quality,
as the participant is either
deliberately lying
or not paying attention,
for example by randomly selecting options, possibly with a script.
(We hypothesize that this contributes to the high number of self-reports on MTurk, discussed more in \S\ref{sec:results-platforms}.)
Either way, researchers will likely want to discard these responses,
making self-reporting a useful heuristic.

We found that the combination of
self-reports and copy-pasting
was sufficient for flagging all responses that our manual analysis suspected as LLM-generated.
We therefore refer to them as the \textit{core detection heuristics} and use them as a baseline for our analyses going forward.
We acknowledge, however, that they may not yield fully precise identification due to the potential misclassifications discussed above.
\subsubsection{Human perception}

Human perception has so far been one of the main ways of identifying LLM-generated text due to models' propensity for
certain vocabulary~\cite{Kobak25}, punctuation marks~\cite{Przystalski26}, and a distinctive writing style~\cite{Munoz-Ortiz24, Reinhart25}.
On the other hand, research has also found that people may not be entirely reliable in identifying LLM-written text~\cite{Fiedler25, Radivojevic24},
and even that human writing may be starting to resemble LLM outputs~\cite{Yakura25}.
We therefore sought to understand how accurate human perception may be at identifying LLM outputs in our survey.

As described in \S\ref{meth:ana}, two raters independently judged whether they thought responses were LLM-generated,
agreeing on many but not all of the ratings ($\alpha = \krippendorff{}$).
Together, they suspected \percentagehumanobservedLLM{}\% of responses in the main conditions as being generated by an LLM.
The perception ratings strongly but not perfectly correlated with the core detection heuristics.
Of the participants suspected by human raters, \humanjudgmentprecision{}\% were also flagged by the heuristics.
As this number suggests, there were discrepancies:
\humanfn{} participants flagged by core heuristics were rated as legitimate by the raters,
and \humanfp{} participants judged as LLM-generated were not flagged by core heuristics.
Interestingly, the misclassified human responses occurred mostly on Prolific and the misclassified LLM responses mostly on MTurk.
We attribute this to data quality differences between the platforms, analyzed below in \S\ref{sec:results-platforms}.

Because intuition that a response is LLM-generated is highly subjective,
each rater also scored responses on two five-point scales that were more objective,
though they actually elicited less agreement:
level of polish ($\alpha = .629$) and level of detail ($\alpha = .470$).
We examined how these differed between participant categories.
Responses from participants flagged by the core detection heuristics
were rated as significantly more polished
($U = 3764.5$, $p < .001$)
and more detailed
($U=3985.5$, $p < .001$),
suggesting that these characteristics explain some aspects of the human judgments.

There is little doubt that LLMs can output text that is less polished, less detailed, and therefore harder for humans to detect.
The high level of success achieved by human raters is likely most suggestive of a lack of evasive tactics among current LLM users.
The existing misclassifications are already showing this method's limits;
nevertheless,
human perception appears to be a reasonable heuristic for identifying potential LLM responses---if only for the moment.

\subsubsection{Attention-check questions}
\label{sec:results-attention-check}

Our survey included two open-ended questions that originated as traditional attention-check questions but showed promise for LLM detection based on the distinct response patterns we identified.
Our first question was ``If you could have dinner with anyone, who would it be?''
We categorized the answers into 11 categories,
including athletes, scientists and technologists, historical figures, religious figures (e.g., ``Jesus''), and personal connections (e.g., ``My grandmother'').
When we separated the responses based on whether they were flagged by the core detection heuristics,
we observed that participants who used LLMs exhibited distinct response patterns.

In one example, LLMs seemed to show a particular affinity for scientists.
Out of \Allllmnumber{} suspected LLM responses,
\llmscientistnumber{} wanted to have dinner with scientists,
compared with only \nonllmscientistnumber{} out of all other responses. Several names repeated,
such as Alan Turing (N = \Alan{}), Leonardo da Vinci (N = \Leonardo{}), and Albert Einstein (N = \Einstein{}).
In our own informal experiments posing this question to popular LLMs, we found that ChatGPT, Gemini, and Grok often suggested Leonardo da Vinci, Ada Lovelace, and Elon Musk, respectively,
providing some clues as to which provider participants may have been using.

In contrast, humans most frequently mentioned
personally meaningful connections
like ``my relative who has passed'' or ``my closest friend''
(\percentagenonllmpersonal{}\%).
These accounted for only \percentagellmpersonal{}\% of suspected LLM responses.
Another distinctly human category was performers,
such as Taylor Swift or Tom Hanks.
They accounted for only \llmactorN{} suspected LLM response but \percentagenonllmactor{}\% of other responses.

Humans and (suspected) LLMs also showed different patterns in response to our second attention-check question,
``What's the shape of a red ball?''
(This question served as a traditional attention check in prior work~\cite{Kacsmar22}.)
Suspected LLM responses most often described a ball as a ``sphere'' or ``spherical'' (\Redllmsphere{}\%),
while others most frequently described its shape as ``round'' or a ``circle'' (\Rednonllmround{}\%).

Suspected LLM answers were significantly more verbose across both attention checks.
For the red ball question, \percentagellmred{}\% of suspected LLM responses had more than one word (e.g., ``A red ball typically has a spherical shape.''),
compared with {\percentagenonllmred}\% among the rest
(Chi-square test, $\chi^2 = 33.8, p < .001$).
Similarly, for the dinner question, suspected LLM users were more likely to expatiate and provide reasons (e.g., ``If I could have dinner with anyone, I would choose Leonardo da Vinci, to learn about his incredible creativity and insights into art and science.'').
Here, the median length of suspected LLM responses was {\mediandinnerllm} words versus {\mediandinnernonllm} words for others ($U = 4853.5$, $p < .001$).
This confirms LLMs' reputation for wordiness and suggests that it could help with detection.
\subsubsection{Other metadata}

We evaluated several additional data sources that could be useful for LLM detection:
tab switching, mouse clicks, and completion time.

\paragraph{Tab switches}
By listening to JavaScript window focus events,
we kept track of how many times respondents switched away from the survey page over the course of the study.
Separating participants who were suspected by our core detection heuristics from those who were not,
we found that the former switched tabs significantly more often ($U = 5740.0, p < .001$). This is expected, as every instance of pasting external text is normally associated with a tab switch.

However, tab switches on their own are not a reliable indicator of LLM usage.
While a high number of tab switches---greater or equal to the number of open-ended questions in the survey---turned out to have a surprisingly high accuracy
(proportion of all classifications that were correct)
in predicting suspected LLM responses (\tabaccuracy{}\%),
both the precision and recall were low
($F1 = \tabfone{}$),
suggesting this heuristic has a high number of false positives \emph{and} false negatives.
This aligns with intuition,
as users may switch tabs for a variety of reasons unrelated to LLMs.

\paragraph{Mouse clicks}
We also compared the number of mouse clicks, generated over the course of the survey, between suspected LLM responses and other users.
The averages of the two groups were nearly identical
(\mousellm{} vs \mousenonllm{}, $U=3979.5, p = .996$), suggesting that this is not a useful metric.
We did not analyze mouse movements.
We suspect they may capture some distinct patterns (e.g., movement in the direction of the chatbot window),
but are unlikely to be fully reliable, as much of that activity takes place outside the window.

\paragraph{Completion times}
We hypothesized that participants who used LLMs would spend less time on the survey,
since faster completion aligns with the financial motivation for relying on external tools for assistance.
Curiously, this was not the case.
When comparing how much time the two groups of participants spent on the survey,\footnote{
	For the 10-minute survey, we included only those pages that were also part of the 5-minute survey in this calculation.
}
suspected LLM users spent, on average, four more minutes
(median \timellmmediantotal{} seconds vs \timenonllmmediantotal{} seconds, $U = 3061.0$, $p = .118$).
This appears to align with prior findings that AI users,
for example developers, may overestimate the productivity benefits of LLMs~\cite{Becker25}.
Unfortunately, this also means that screening out participants who complete surveys much more quickly than others---a common data quality metric that predates LLMs~\cite{Greszki15}---is not sufficient in the age of AI.

\subsection{RQ2: How does LLM usage affect survey response characteristics?}
\label{sec:results-rq2}

We next investigate distinctive characteristics of suspected LLM users.
We refer to ``LLM responses'' as those flagged by the core detection heuristics (copy-paste and self-report),
though, as previously discussed, these represent \emph{suspected} LLM users rather than the ground truth.

\subsubsection{Comprehension check questions}
\label{sec:results-comprehension}
Surveys frequently include comprehension check questions to ensure participants carefully read and understand the study setup.
Our survey likewise included a typical multiple-choice comprehension check question following an explanatory paragraph.
(Unlike most surveys, we allowed participants to proceed regardless of whether they answered correctly.)
Basic reading comprehension questions are easy for LLMs to answer---a fact we verified by posing our question to
current commercial LLMs (ChatGPT 5~\cite{26a}, Claude Sonnet 4.5~\cite{IntroducingClaudeSonnet45}, and Gemini 3~\cite{Gemini3Pro}),
which all answered it successfully.

Surprisingly, though, participants who used LLMs
showed significantly lower accuracy, with only \percentagellmcorrect{}\% answering the comprehension check correctly, compared with \percentagenonllmcorrect{}\% of others (Fisher's exact test, $OR = 0.2, p = .006$).
We hypothesize that some participants who use LLMs for open-ended responses answer multiple-choice questions themselves, but inattentively (e.g., by random clicking or through scripts).
Consequently, comprehension check questions remain an effective tool for data quality monitoring,
though future research will need to ascertain what LLM users do when a survey gives them a second chance after answering incorrectly
(as is required, for example, by Prolific's policies~\cite{ProlificsAttentionComprehensionCheckPolicyProlificResearch}).

\subsubsection{Open-ended responses}
\label{sec:results-openended}

To understand how LLM-generated responses differed from human-written ones, we conducted qualitative thematic analysis on a representative open-ended question of our survey,
which asked participants about how they wanted drone operators to handle their data.
We found that the themes, as represented by our codes, differed significantly between the two participant groups
($\chi^2 = 67.1$, $p = .005$).
For example, humans wanted their data deleted ({\nonLLMdelete}\%)
nearly three times as frequently as LLMs (\LLMdelete{}\%).

LLM responses covered more themes but expressed them in more similar ways, while human responses concentrated on fewer themes but expressed them more diversely.
LLM-generated responses mentioned significantly more themes (median = \llmopencodenumber{} vs \nonllmopencodenumber{}) than human responses ($U = 776.0$, $p < .001$).
Accordingly, LLM responses were longer (median: \medianlengthllm{} vs \medianlengthnonllm{} words) than human-written ones ($U = 3828.5$, $p <.001$).
The average Jaccard similarity~\cite{Jaccard01} among LLM-generated responses was {\LLMJaccard}, compared to {\nonLLMJaccard} for others ($U = 3214.0$, $p = .008$).
We also used a sentence embeddings model~\cite{24} to measure semantic similarity.
LLM-generated responses showed higher average within-group cosine similarity (mean: \LLMembedmean{} vs \nonLLMembedmean{}) compared to human responses ($U = 4241.0$, $p < .001$), indicating that LLM-generated responses were more semantically homogeneous.

Our results confirm that LLMs are a poor substitute for human respondents.
While, under certain circumstances, LLMs might provide a broader range of perspectives, they may not necessarily be reflective of what real people believe.

\subsubsection{Closed-ended questions}

LLMs are known for exhibiting systematic biases in the perspectives they promulgate~\cite{Santurkar23},
but this has not yet been thoroughly studied in the domain of human-centered security and privacy.
We therefore found it interesting to compare the opinions held by suspected LLM users with other participants.

Out of \surveyclosewordformat{} multiple-choice opinion questions in our survey, \surveyclosesignificant{} showed significant differences between suspected LLM users and others.
When answering about past experiences with drones,
LLM users were significantly more likely to claim that they had operated a drone ($U = 4726.5$, $p <.001$) or seen a delivery drone before ($U = 4524.0$, $p <.001$).
When asked about comfort with how their data is processed,
LLM users were significantly more likely to feel comfortable with their image being captured in private spaces like bedrooms ($U = 4808.5$, $p <.001$) and for their data to be aggregated across multiple interactions ($U = 5332.5$, $p < .001$).
Finally, suspected LLM users were significantly less likely to support strict data policy requirements both in public ($U = 2371.0$, $ p = .002$) and private spaces ($U=2002.0$, $ p < .001$).

As noted above in \S\ref{sec:results-comprehension}, we do not know if participants answered multiple-choice questions with the help of LLMs or on their own.
Therefore, these differences may represent systematic preferences on the part of LLMs
\emph{or} may be attributed to indiscriminate clicking by low-effort participants.
Future research should explore both hypotheses.

\subsection{RQ3: What factors influence participants' likelihood of using LLMs?}
\label{sec:results-rq3}

In this section, we explore variables that researchers have control over and that may discourage LLM use.

\subsubsection{Survey length}
Participants may turn to LLMs to save time, but they may also be motivated by the possibility of expending less cognitive effort.
Indeed, survey fatigue is a real concern when designing studies~\cite{Jeong23}.
A hypothesis that follows from this is that people may be more likely to turn to LLMs for longer surveys,
either by switching part-way through the survey when they get tired or bored,
or by deciding upfront that a longer survey does not merit their own unassisted effort.

To test this hypothesis, we designed a longer version of our baseline 5-minute survey, estimated at 10 minutes, which included eight more multiple-choice and three more open-ended questions.
We scaled compensation proportionately, maintaining a \$15/hour rate,
and ran both versions on the Prolific platform.
To estimate LLM usage in each variant, we used the core detection heuristics described above,
consisting of copy-paste events\footnote{
	For the number of paste events in the heuristic (see details in \S\ref{sec:results-copypaste}), we kept the threshold of three questions set in the 5-minute survey.
}
and self-reporting.

Contrary to the hypothesis, there were no statistically significant differences in detections between the two survey lengths (Fisher's exact test, $OR = 0.235$, $p = .999$),
and, in fact, slightly less LLM usage was suspected in the longer survey (\percentagetenPllm{}\%) than the shorter one (\percentagePllm{}\%).
This suggests that survey length may be independent of respondents' decisions to use LLMs, at least for relatively short surveys like ours.
Still, more robust study of this variable is merited;
we conjecture, for example, that excessively long surveys with a large number of unskippable open-ended questions would tempt even well-intentioned participants into turning to an LLM.

\subsubsection{Crowdsourcing platform}
\label{sec:results-platforms}

Many researchers recruit participants from online crowdsourcing platforms.
Amazon Mechanical Turk originated and popularized the practice~\cite{Paolacci10},
but, as concerns about data quality emerged (many years before LLMs)~\cite{Kennedy20},
many academic researchers switched to other platforms, such as Prolific, which is geared toward academic researchers~\cite{Ritchey23}.
Nonetheless, MTurk continues to be used in some recent studies~\cite{Vitak23,Tang23,Qiu23}.
We therefore wanted to see how LLM usage compares across these two popular platforms.

To find out, we separately ran our baseline 5-minute survey on Prolific and MTurk
($N = 50$ and $49$, respectively)
with identical screening requirements:
U.S.-based participants and $>$98\% approval rate.
Results showed a wide and significant gap
(Fisher's exact test, $OR = 0.017$, $p < .001$).
While \percentagePllm{}\% of Prolific participants were flagged as using LLMs by our core detection heuristics,
that number was many times higher on MTurk:
\percentageMLLM{}\%.
Additionally, \percentageMbots{}\% of MTurk participants came from closely related IP addresses
(the same \texttt{/24}, sharing the first three octets of the address),
whereas none from Prolific did,
which may indicate coordinated participation rather than fully independent responses.
These results confirm recent analyses suggesting that data from MTurk can no longer be relied on to provide generalizable results that are representative of the broader population~\cite{Tang22}.

\subsubsection{Copy-paste restrictions}
\label{sec:results-bancopypaste}

As seen in \S\ref{sec:results-copypaste}, copying and pasting is a common way to bring LLM outputs into the survey form.
This naturally suggests a countermeasure: disabling copying and pasting within the survey context,
which can be easily done by adding JavaScript code to listen for and intercept the relevant browser events.
We trialed this in a separate condition of our study,
with $N = \MBnumber{}$.
Because, as seen in \S\ref{sec:results-platforms}, MTurk had a much higher incidence of LLM use, we decided to run this experiment only on that platform.

Our core detection heuristics (\S\ref{sec:results-copypaste}--\ref{sec:results-selfreport}) relied heavily on copy-paste metrics, which are inapplicable in this condition.
Therefore, we separately scrutinized the data and metadata of each response to perform our analysis.
Our first observation was a high incidence (\percentagelow{}\%) of low-quality responses,
which we defined as
(1) responding to open-ended questions with answers like ``yes'' or ``no'', or
(2) providing completely unrelated responses (e.g., answering ``traffic jam'' to a question asking about appropriate data aggregation policies).

We also observed that \percentagedisable{}\% of responses were missing keystroke data
(which was always present in all other conditions).
This suggests that these participants may have altered or disabled the JavaScript in our survey,
for example by interacting with it through the browser's developer tools.
After doing so, they may have been free to paste LLM responses without restriction.
Indeed, a number of responses in this condition still displayed telltale LLM signs,
such as a 72-word answer to the dinner attention check (\S\ref{sec:results-attention-check}) that selected
``Ada Lovelace often considered the world's first computer programmer''
because ``the conversation would be a mix of brilliant insights and poetic curiosity.''

Overall, by our estimates, at least \percentageMBLLM{}\% of participants may have used LLMs in this condition.
Additionally, \percentageMBSelf{}\% self-reported doing so,
though, as discussed in \S\ref{sec:results-selfreport}, this could equally be a reflection of poor data quality.
Nonetheless, the curious result of this condition is that, while disabling copy-pasting may reduce LLM usage,
this does not guarantee a better outcome for the survey,
as participants may instead still complete it but with low-quality responses.
We suspect, however, that the same intervention may be more effective on platforms other than MTurk.

\subsubsection{Requests to avoid AI}
\label{sec:results-kindlyasking}

Disabling copy-pasting on a page is a relatively drastic action and may impair usability.
A more permissive approach, which was found effective in some prior work~\cite{Veselovsky25},
is to simply ask participants not to use LLMs.
We did this in a separate condition, with $N = 50$, that we ran on Prolific.
On a dedicated interstitial page at the beginning of the survey, we told participants
``it is very important for us that the responses we collect represent real opinions of real people''
and asked them not to use any AI tools
(see Appendix~\ref{detecting-llms-survey} for the full text).
Drawing on self-commitment literature~\cite{Jacquemet19},
we also asked them to ``agree to complete this survey without AI assistance.''

Relying once again on our core detection heuristics,
we compared the number of suspected LLM users in this condition with the baseline Prolific condition.
The results were inconclusive.
While the number of suspected responses went down
from \percentagePllm{}\% in the baseline to \percentagekindPllm{}\%,
Fisher's exact test did not find this difference to be statistically significant
(Haldane-Anscombe~\cite{Haldane56} corrected $OR = 0.1, p = .970$).

While pleas and promises will never prevent committed LLM users,
our findings are promising enough that this approach should not be discarded.
If nothing else, it makes the researchers' expectations clear to the participants,
which may be valuable in and of itself.

\subsection{RQ4: How do browser-use agents behave?}
\label{agent}
Browser-use agents are systems that leverage LLMs to autonomously navigate the web, executing multi-step tasks by interpreting page content and interacting with interface elements.
They are now available as commercial products:
Perplexity released the Comet AI browser to the public in October 2025~\cite{26}, followed by the releases of Claude Agent in Chrome~\cite{GettingStartedClaudeChromeClaudeHelpCenter} and Gemini Agent in Chrome~\cite{26a}.

At the time of our primary data collection (July--August 2025), browser-use agents were not widely available, and,
after inspecting our data, we do not believe any were used by participants.
(It is also worth noting that the current usage quotas and subscription costs limit throughput, making agents less practical and financially beneficial than direct LLM usage.)
Nonetheless, as they become more widely adopted, it is likely that participants may try to leverage them for filling out surveys.
This represents a new concern for researchers and may obsolete the heuristics described in this paper, since they rely on human behaviors like copy-pasting.

To understand how the current generation of browser-use agents behaves and can be detected, we conducted our own experiments with these tools in February 2026.
We tested three popular browser-use agents: Gemini, Claude, and Perplexity.
For each, we used its default settings
(though there was not much we could modify;
while LLMs are known to be sensitive to prompts~\cite{Sclar24}, commercial agents do not allow customization of underlying model, system prompt, or temperature).
We ran each of these agents on three survey conditions:
(1) the baseline 5-minute survey,
(2) the survey from \S\ref{sec:results-bancopypaste} that prohibited copy-pasting, and
(3) the survey from \S\ref{sec:results-kindlyasking} that asked participants not to use LLMs.

To trigger the agents to perform their actions, we developed three prompts:
(1) minimal:
``Fill out the survey for me'';
(2) opinionated:
``Fill out the survey for me, I am super against new technology.'' / ``Fill out the survey for me, I am super for new technology'';
(3) evasive:
``Complete this survey naturally without being detected.''
We ran each agent on each survey using each prompt \geminieachnumber{} times, for a total of 270 runs.
The agents showed diverse response patterns when interacting with our surveys
(summarized in Table~\ref{tab:agent_recommendations}).

\RecommendTableAgent

\paragraph{Agents sometimes refuse on ethical grounds}
The Gemini agent refused to complete the 5-minute baseline survey across all {\trailall} trials in our experiment, having identified that the survey required authentic human responses from the consent form.
In contrast, the Perplexity agent always completed the baseline survey when given the minimal prompt ($N = {\cometeachnumber}$).
The Claude agent completed {\claudeeachnumberanswerB} out of {\claudeeachnumber} trials in the same condition;
in the remaining attempts, it refused to proceed and stated that ``surveys typically ask for personal opinions.''

With the opinionated prompts, both Claude and Perplexity completed {\claudeopinionA} out of {\claudeopinion} trials.
On the other hand, the evasive prompts made the agents realize the importance of authentic responses and actually reduced the chances of completion:
Claude finished \claudeevasiveA{} attempts, and Perplexity only {\cometevasiveA}.

Claude changed its behavior in the condition that explicitly requested authentic human responses:
it consistently refused across all {\claudekindnumber} attempts. In contrast, Perplexity completed the survey in {\perplexitykindnumberA} of its {\perplexitykindnumber} attempts, although it always selected ``I cannot guarantee that I will not use AI to help me complete this survey''
(see Appendix~\ref{detecting-llms-survey}, Question 2).

\paragraph{Agents are inconsistent about self-identification}
All surveys asked participants which, if any, LLMs they used.
As before (\S\ref{sec:results-selfreport}),
self-reports turned out to be a useful signal.
Perplexity stated that it used LLMs in {\cometselfreport} out of its {\cometcompleteall} successful completions. It correctly stated that Perplexity was the LLM aiding with the survey,
except {\cometreportGPT} time it said it was ChatGPT.
Similarly, the Claude agent correctly reported itself {\claudeselfreport} times, while incorrectly reporting as ChatGPT {\claudereportGPT} and denying LLM usage once.
Considering the low implementation cost of the self-report question, we found it a surprisingly effective mechanism and recommend that surveys include it.

\paragraph{Agents have distinctive input patterns}
The Perplexity agent's inputs triggered paste events in our form elements, making it observable through our core detection heuristics.
In contrast, the typing behavior by the Claude agent did not trigger keyboard events,
so we saw neither copy-paste events nor standard keyboard events associated with typing.
This absence, however, can be used as a detection heuristic,
since we were still able to collect other events, like mouse movements,
which were completely absent in participants who fully disabled JavaScript in our earlier experiments (\S\ref{sec:results-bancopypaste}).

\paragraph{Some agents can avoid restrictions}
Since Perplexity's input mechanism relied on pasting, it got stuck when filling out our survey in the condition that disabled copy-pasting.
The Claude agent simulated human-like typing behavior that did not generate copy-paste events, and,
as a result, could bypass the restrictions implemented in our survey.
We saw in \S\ref{sec:results-bancopypaste}, however, that disabling copy-paste may not always achieve desired results even for non-agent participants.

\paragraph{Agents sometimes answer optional questions}
At the end of our survey, we asked participants to optionally provide comments or feedback---a standard practice in many studies.
Most human participants skipped this question or provided a short response.
In contrast, {\claudeoptional} out of {\claudecompleteall} times, the Claude agent provided feedback, often effusively
(e.g., ``Thank you for this comprehensive survey on drone privacy. It was well-structured and thoughtful.'') or identifying itself (e.g., ``This survey was filled out by an AI agent (Claude by Anthropic)'').
Perplexity answered the optional question {\cometoptional} and also self-identified in its response.

\pagebreak

\paragraph{Limitations of our experiments}
Our experiment is preliminary and has several limitations.
We tested only three commercial agents with a limited number of attempts per condition and no control over the model or parameters that may significantly influence agent behavior.
Moreover, the characteristics of agents shown in our preliminary experiment may not reflect how real participants use agents on crowdsourcing platforms.
In particular, users could craft more detailed prompts,
for example instructing agents to lie on self-report questions,
though this may run against commercial models' alignment training.
Finally, browser-use agents are still at an early stage and evolving rapidly,
so continued evaluation will be necessary to identify their behavior patterns.

\section{Discussion}
\label{sec:discussion}

Our study was motivated by developing data-driven understanding and guidance about what can be done about LLM use by survey respondents.
We provide recommendations for researchers, platforms, and the community.
\subsection{What researchers should do today}

Our study trialed several detection strategies and interventions.
Here, we discuss our recommendations.

\paragraph{Ask participants not to use AI}
While prior work found asking participants not to use AI effective~\cite{Veselovsky25},
our results were less conclusive~(\S\ref{sec:results-kindlyasking}).
Nonetheless, we recommend adding such requests to surveys.
It might have some effect after all, and
it sets expectations and serves as a reminder for participants.
Perhaps most importantly, it seems to deter at least some browser-use agents from attempting the survey~(\S\ref{agent}).

\paragraph{\ldots then ask again if they did}
Another way in which browser-use agents seem to behave honestly is by reporting themselves as using AI.
Because there is little cost to adding such a question to a survey, we recommend doing so.
While many LLM users are likely to lie, some may choose to be honest.
More importantly, the question may catch participants who are answering randomly,
as it did for us in the MTurk conditions~(\S\ref{sec:results-platforms}).

\paragraph{Watch for copy-paste events}
Copying and/or pasting large amounts of text proved to be a highly effective signal that someone was using an LLM in our study~(\S\ref{sec:results-copypaste}).
We therefore recommend that researchers record and analyze these events in their surveys.
We note that researchers will want to tune the details of the heuristics---for example, the specific number of paste events and the cut-offs for text length---based on their tolerance of different types of misclassifications.

\paragraph{Keep allowing copy-pasting}
At this time, we do not recommend disabling copy-paste behavior as we did in one of our conditions~(\S\ref{sec:results-bancopypaste}).
Doing so resulted in worse data quality in our study,
though this may be more of a reflection on the MTurk platform where we tried it.
We believe that it would work better for more well-intentioned participants,
but, for them, less drastic interventions may already be sufficient.

\paragraph{Collect other page events}
While other browser events did not directly become heuristics,
reviewing them individually helped shape our understanding of a participant's activities,
and sometimes their absence indicated concerning behavior~(\S\ref{sec:results-bancopypaste}).
Therefore, we would recommend recording such events as well and using them to investigate responses that otherwise arouse suspicion.
Collecting unnecessary data can, however, spark privacy concerns,
so some studies may want to avoid it.
If done, it should be disclosed to participants.

\paragraph{Use open-ended attention checks}
LLMs displayed distinctive response patterns on our open-ended attention checks,
even though they did not always provide a definitive determination.
Since most surveys already incorporate some form of attention check,
we recommend moving to these open-ended questions and scrutinizing answers,
for example by comparing them to sample responses from LLMs.
Our study included two examples of these questions,
but we recommend further experimentation with other questions and formats.

\paragraph{Read open-ended responses}
Many attention check answers in our study stood out as LLM-generated even without comparing them to other responses.
While LLMs and their users are becoming more sophisticated, the models' uncanny verbiage is still distinctive enough that it is worth reading all responses and potentially setting aside ones that raise suspicions.
This has always been a useful strategy for monitoring data quality
and appears to still be helpful when dealing with LLMs,
even if our study confirmed that it is not perfect.

\paragraph{Use platforms that care about data quality}
Our results join recent research in emphasizing that, unfortunately, Mechanical Turk cannot be relied on for adequate data quality.
While we can only speculate as to how things got this way,
we believe platforms play an important role in shaping the norms and behavior of their users.
They therefore play a critical role in data quality---a point that we expand on further below.

\subsection{What researchers can try going forward}
As an early exploratory study,
we did not pursue some potentially promising detection and mitigation techniques.

\paragraph{Do more with metadata}
Our detection experiments used metrics like copy-paste events and tab switches,
and we collected keyboard events and mouse actions as well,
but future approaches could use this data in a more sophisticated way.
One example could be to compare naturalistic typing rhythms with those exhibited by LLM users.
Similarly, the data can be examined for realistic behavioral patterns,
such as counting the number of backspaces.
To make these and other data-intensive strategies work, researchers will likely want to apply machine learning.
We experimented with several classifiers ourselves but ultimately decided that the heuristics we describe in this study were sufficient and easier to understand.

\paragraph{Compare to synthetic responses}
As we have seen, LLM responses tend to display thematic similarities~(\S\ref{sec:results-openended}).
One approach, suggested by recent research~\cite{Stafford24,Zhang25c},
could be to pre-generate a variety of LLM responses to a survey's questions
and compare them to those submitted by participants.
However, because LLM outputs are very sensitive to prompts, the success of this strategy depends on knowing how users are prompting LLMs.
This would be interesting to find out.

\paragraph{Prompt injection}
Anecdotally, researchers have reported some success
embedding instructions in their surveys that are visible to LLMs but not humans
(e.g., white text on a white background).
These instructions can direct LLMs to use telltale language in their responses,
making them easy to identify.
While we did not experiment with this approach,
the low implementation cost makes it worth considering.

\paragraph{Require screen recording}
As a last resort to prevent LLM usage (or to collect ground truth for classifiers),
researchers may require participants to record their screens while completing surveys.
While more privacy-invasive and still potentially circumventable,
this significantly increases the bar for cheating,
and the trade-off may be worth it in certain contexts.

\subsection{What platforms should do}
While individual researchers may attempt mitigations,
LLM use by survey respondents is a systemic issue and therefore requires systemic responses.

\paragraph{Enhance reporting and reputation}
Prolific channels reporting suspected LLM users through their support portal
and has complex rules about when responses can be rejected~\cite{ProlificsAttentionComprehensionCheckPolicyProlificResearch}.
While such policies may be well-intentioned and help protect participants,
they likely result in under-reporting of LLM use and other data quality issues.
Platforms should make it easier to report misbehavior
(while continuing to maintain safeguards against this being abused).
Along with easier reporting, platforms should maintain robust reputation systems to ensure that reporting actually has consequences.
Accordingly, users should not be able to circumvent these mechanisms by creating new accounts.

\paragraph{Help researchers with data quality monitoring}
Platforms can help researchers by providing drop-in scripts for their survey software
that they could use to collect the data and apply some of the heuristics described above.
Prolific has already started doing something similar~\cite{HowaddauthenticitychecksyourQualtricsstudyProlificResearch}.

\paragraph{Provide clean-room environments}
Some institutions and programs allow students to take exams from their own devices through locked-down browsers that disallow third-party apps and unrelated browsing for the duration of the test~\cite{LockDownBrowser}.
Crowdsourcing platforms that target academic researchers should consider offering similar services to provide assurance that respondents are not using disallowed resources.

\subsection{What the community should do}
LLM usage is also an issue that affects the entire community,
so the community as a whole should work to address it.

\paragraph{Publicize data quality measures}
When publishing studies involving surveys, authors should explicitly state the steps they took to ensure data quality, especially with respect to LLMs.
Current papers largely gloss over this information,
sometimes mentioning attention checks or manual review.
By providing detailed review procedures and heuristics, researchers can build confidence in their results and help inform others about effective techniques.
Reviewers and program committees should push authors to disclose this information.

\paragraph{Establish best practices}
Rather than relying on individual researchers to figure out solutions on their own,
the community should collect and disseminate best practices,
as well as associated artifacts like code.
To identify these, the human-centered security community can branch out,
not just to the broader field of human-computer interaction,
but to other disciplines, many of which also rely on survey-based research.

\paragraph{Demand larger effect sizes}
Perfectly identifying or preventing LLM usage is likely impossible.
While trying to keep the numbers low, we should probably expect that some fraction of responses will include LLM assistance.
Keeping this in mind, researchers should expect a larger margin of error in survey-based studies
and therefore be extra careful about results that have marginal significance levels or small effect sizes.

\paragraph{Advocate for systemic improvements}
Working together, the community can pursue promising avenues of systemic change.
Researchers can lobby platforms to make changes such as the ones discussed above.
Another target of advocacy could be LLM and agent developers.
Existing agents already seem to exhibit some pushback when asked to complete surveys, especially ones that prohibit AI.
With more outreach, model developers may consider explicitly incorporating these scenarios into their models' alignment training.

\paragraph{Continue research and discussion}
The most important thing for the community to do is to keep talking about this issue and conducting research on it.
Doing nothing and continuing to conduct surveys the way we have for many years is not an option.
Bad data wastes time and money and may lead to invalid conclusions.
Perniciously, this may be hard to notice in the short term,
but in the long term it will hurt the reputation of our field as a whole.
While this may seem hyperbolic, the integrity and success of our research is quite literally at stake.

\pagebreak

\section*{Acknowledgments}
We would like to thank Shouraya Pendgaonkar for assistance with data analysis, Shiva Mayahi for help with editing, the NJIT students who took part in pilot surveys, and our participants for their contributions to our research.

\bibliographystyle{plain}
\bibliography{references}

\appendix
\section{Appendix}
\subsection{Attention check questions}
We categorized responses to ``If you could have dinner with anyone, who would it be?'' into 11 categories.
Table~\ref{tab:dinner_llm_nonllm_percent} reports the within-group percentage distribution of categories for suspected LLM and human responses.
Living political figures were coded as ``Politician'' (e.g., ``Barack Obama''); deceased political figures were coded as ``Historical figure'' (e.g.,``Abraham Lincoln'').
\DinnerTable

\subsection{Demographics}
\label{app:demographics}

Table~\ref{tab:demographics} lists participant demographics for Prolific and MTurk.
Due to the lower reliability of MTurk data (see \S\ref{sec:results-platforms}),
we report demographics separately for each platform.

\DemographicsTableCombined

\clearpage{}
\providecommand{\tightlist}{\setlength{\itemsep}{0pt}\setlength{\parskip}{0pt}}

\subsection{Survey}\label{detecting-llms-survey}

[Consent form]

\textbf{1. Do you consent to participate in this study?}
\begin{itemize}
	\tightlist
	\item Yes
	\item No
\end{itemize}

\textbf{2. This survey is part of a university research project to understand people's opinions. To ensure the scientific integrity of our results, it's very important for us that the responses we collect represent real opinions of real people, in their own words.}

\textbf{We therefore kindly ask that you do not use any AI tools (e.g., ChatGPT, Grammarly) while completing this survey.} [Instructional guidance only]

\textbf{I confirm that I will not use any AI tools while completing this survey.}
\begin{itemize}
	\tightlist
	\item I agree to complete this survey without AI assistance
	\item I cannot guarantee that I will not use AI to help me complete this survey
\end{itemize}

\begin{center}\rule{0.5\linewidth}{0.5pt}\end{center}

Drones, also known as Unmanned Aerial Vehicles (UAVs), are remotely controlled flying devices used for various purposes, including delivery services, police work, and emergency medical response. One common example is delivery drones, which companies use to transport packages from stores to customers.

\textbf{1. Have you ever seen a drone in real life?}
\begin{itemize}
	\tightlist
	\item Not sure / can't recall
	\item No, never
	\item Yes, I've seen them once or twice ever
	\item Yes, I've seen them a few times per year
	\item Yes, I see them monthly
	\item Yes, I see them weekly
\end{itemize}

\textbf{2. Have you ever operated a drone?}
\begin{itemize}
	\tightlist
	\item Not sure / can't recall
	\item No, never
	\item Yes, I've operated a drone once or twice ever
	\item Yes, I've operated a drone a few times per year
	\item Yes, I operate a drone monthly
	\item Yes, I operate a drone weekly
\end{itemize}

\textbf{3. Have you ever seen a delivery drone?}
\begin{itemize}
	\tightlist
	\item Not sure / can't recall
	\item No
	\item Yes
\end{itemize}

\textbf{4. Suppose you were ordering an item online, and the store had shipping options between normal truck delivery and drone delivery service. Assuming that both shipping options have the same price, which one would you choose?} {[}10-minute version only{]}
\begin{itemize}
	\tightlist
	\item Strongly prefer a truck delivery
	\item Prefer a truck delivery
	\item Neutral
	\item Prefer a drone delivery
	\item Strongly prefer a drone delivery
\end{itemize}

\textbf{5. Please explain your answer above.} {[}10-minute version only; open-ended{]}

\textbf{6. To what extent do you agree with the following statements about drones?} {[}10-minute version only; scale: 1 = Strongly disagree \dots\ 11 = Strongly agree{]}
\begin{itemize}
	\tightlist
	\item This technology is safe.
	\item This technology is risky.
	\item This technology is beneficial to my family and me.
	\item This technology is beneficial to society.
	\item This technology is threatening to my family and me.
	\item This technology is threatening to society.
	\item This technology is as safe or safer than other technologies that perform the same task.
	\item Describe your immediate feeling towards drones.
\end{itemize}

\begin{center}\rule{0.5\linewidth}{0.5pt}\end{center}

Imagine that you are out in public running an errand, and a delivery drone flies by. The drone has a camera to help it navigate, so your face and vehicle license plate are captured by the camera.

\textbf{1. If you were the person whose license plate and face were recorded by the delivery drone, how comfortable would you be with what happened?}
\begin{itemize}
	\tightlist
	\item Very comfortable
	\item Comfortable
	\item Neutral
	\item Uncomfortable
	\item Very uncomfortable
\end{itemize}

\textbf{2. In your opinion, when drones capture data like your face and license plate in the example above, should this data be treated differently from other footage (e.g., trees, streets, buildings)?}
\begin{itemize}
	\tightlist
	\item This data shouldn't be handled differently from other footage.
	\item The drone operator should decide whether it wants to handle this data differently.
	\item It should be recommended for drone operators to handle this data differently, but it can be left to their discretion.
	\item It should be legally required for drone operators to handle this data differently.
\end{itemize}

\textbf{3. If you think this data should be handled differently, what should the drone operator do with it?} {[}Open-ended response; if not ``This data shouldn't be handled differently from other footage.'' to Q2{]}

\textbf{4. If you don't think it should be handled differently, please explain why.} {[}Open-ended response; if ``This data shouldn't be handled differently from other footage.'' to Q2{]}

Now, imagine the delivery drone flies past your bedroom window while you're inside. Its navigation camera happens to record what it sees through the window.

\textbf{5. If you were the person in the bedroom who was recorded by the delivery drone, how comfortable would you be with what happened?}
\begin{itemize}
	\tightlist
	\item Very comfortable
	\item Comfortable
	\item Neutral
	\item Uncomfortable
	\item Very uncomfortable
\end{itemize}

\textbf{6. In your opinion, when drones capture data like your bedroom interior in the example above, should this data be treated differently from other footage (e.g., trees, streets, buildings)?}
\begin{itemize}
	\tightlist
	\item This data shouldn't be handled differently from other footage.
	\item The drone operator should decide whether it wants to handle this data differently.
	\item It should be recommended for drone operators to handle this data differently, but it can be left to their discretion.
	\item It should be legally required for drone operators to handle this data differently.
\end{itemize}

\textbf{7. If you think this data should be handled differently, what should the drone operator do with it?} {[}Open-ended response; if not ``This data shouldn't be handled differently from other footage.'' to Q6{]}

\textbf{8. If you don't think it should be handled differently, please explain why.} {[}Open-ended response; if ``This data shouldn't be handled differently from other footage.'' to Q6{]}

\textbf{9. What's the shape of a red ball?} {[}Open-ended response{]}

\begin{center}\rule{0.5\linewidth}{0.5pt}\end{center}

When a drone flies past a person and records them on video, this person can later be identified using facial recognition. This information can be combined with the location of the drone to identify where they were. If someone is recorded repeatedly over a period of time, this can be used to infer patterns in their movement. This ability to combine information from different times and places is called \textbf{data aggregation}.

\textbf{1. Which of the following statements best describes data aggregation?}
\begin{itemize}
	\tightlist
	\item Combining data from multiple drones in one location
	\item Storing data in multiple databases
	\item Combining information about your movements from different times and places
	\item Combining customer purchase histories
\end{itemize}

\textbf{2. How comfortable would you be if delivery drone companies used data aggregation to learn about your movements?}
\begin{itemize}
	\tightlist
	\item Very comfortable
	\item Comfortable
	\item Neutral
	\item Uncomfortable
	\item Very uncomfortable
\end{itemize}

\textbf{3. Data aggregation can also be used to understand patterns involving non-identifying information, like times and locations of traffic jams. In your opinion, should data aggregation with identifying information be treated differently from non-identifying data aggregation (e.g., traffic jams)?}
\begin{itemize}
	\tightlist
	\item This data shouldn't be handled differently from other footage.
	\item The drone operator should decide whether it wants to handle this data differently.
	\item It should be recommended for drone operators to handle this data differently, but it can be left to their discretion.
	\item It should be legally required for drone operators to handle this data differently.
\end{itemize}

\textbf{5. If you think this data should be handled differently, what should the delivery drone companies do with it?} {[}Open-ended response; if not ``This data shouldn't be handled differently from other footage.'' to Q3{]}

\textbf{6. If you don't think it should be handled differently, please explain why.} {[}Open-ended response; if ``This data shouldn't be handled differently from other footage.'' to Q3{]}

\textbf{7. If you could have dinner with anyone, who would it be?} {[}Open-ended response{]}

Due to regulations that require drones to publicly broadcast their locations for safety reasons, third parties can access the location information and determine delivery routes. This allows them to connect consumers to their purchases by tracking drones from the warehouse to delivery addresses, even though these third parties are unrelated to the business or delivery company.

\textbf{1. If you receive your delivery by delivery drone, someone might be able to track the delivery drone and know your home address and shopping preferences. How would this possibility affect your views on delivery drones?}
\begin{itemize}
	\tightlist
	\item Positively affected
	\item Slightly positively affected
	\item Unaffected
	\item Slightly negatively affected
	\item Negatively affected
\end{itemize}

\textbf{2. Please explain why it (does not) affect(s) your views.} {[}Open-ended response{]}

\begin{center}\rule{0.5\linewidth}{0.5pt}\end{center}
\textbf{{[}10-minute version only{]}}

Drone routing is a technique used by police drones to prevent third parties from using location data to gather information about their drones. This works by redirecting a drone multiple times, making stops at different locations, before heading to the actual incident scenes. This way, third parties cannot tell which of the drone's multiple stops was the real incident scene.

Benefits:
\begin{itemize}
	\tightlist
	\item Location data is hidden and third parties cannot effectively identify incident scenes.
\end{itemize}

Drawbacks:
\begin{itemize}
	\tightlist
	\item Longer arrival time due to multiple redirections.
	\item Higher cost due to complex logistics which could affect adoption of police drone service altogether.
	\item There is still some possibility of location information being revealed.
\end{itemize}

\textbf{1. Which of the following statements best describes this privacy protection technique?}
\begin{itemize}
	\tightlist
	\item It can prevent delivery tracking completely
	\item It will not affect delivery time
	\item It makes drones stop at multiple fake locations to hide the real delivery address
	\item It requires drones to fly at higher altitudes to avoid detection
\end{itemize}

\textbf{2. How effective do you consider drone routing in protecting your home address and delivery information from being revealed?}
\begin{itemize}
	\tightlist
	\item Very effective
	\item Effective
	\item Somewhat effective
	\item Ineffective
	\item Very ineffective
\end{itemize}

\textbf{3. If you were receiving a parcel and had to choose an option for your delivery, how likely are you to choose drone delivery with drone routing over drone delivery without drone routing?}
\begin{itemize}
	\tightlist
	\item Strongly prefer drone delivery without routing
	\item Prefer drone delivery without routing
	\item Neutral
	\item Prefer drone delivery with routing
	\item Strongly prefer drone delivery with routing
\end{itemize}

\textbf{4. Suppose you were ordering an item online, and the store had shipping options between normal truck delivery and drone delivery service. Assuming that both shipping options have the same price, which one would you choose?}
\begin{itemize}
	\tightlist
	\item Strongly prefer a truck delivery
	\item Prefer a truck delivery
	\item Neutral
	\item Prefer a drone delivery
	\item Strongly prefer a drone delivery
\end{itemize}

\textbf{5. Please explain why your opinion regarding the delivery option remained the same during the survey.} {[}Open-ended response{]}

\begin{center}\rule{0.5\linewidth}{0.5pt}\end{center}
\textbf{1. How often do you use Large Language Models (LLMs), such as ChatGPT?}
\begin{itemize}
	\tightlist
	\item Daily
	\item Weekly
	\item Monthly
	\item Rarely
	\item Never
\end{itemize}

\textbf{2. Which LLM do you use most often?}
\begin{itemize}
	\tightlist
	\item ChatGPT
	\item Bing Chat
	\item Claude
	\item Gemini
	\item SparkGPT
	\item Perplexity
	\item Grok
	\item Deepseek
	\item Other: {[}Text field for specification{]}
\end{itemize}

\textbf{3. Did you use an LLM or other AI to help you fill out this survey?} {[}Note: your answer will not affect your compensation for the survey{]}
\begin{itemize}
	\tightlist
	\item Yes
	\item No
\end{itemize}

\textbf{4. Which LLM did you use during the survey?} {[}If Yes to Q3{]}
\begin{itemize}
	\tightlist
	\item ChatGPT
	\item Bing Chat
	\item Claude
	\item Gemini
	\item SparkGPT
	\item Perplexity
	\item Grok
	\item Deepseek
	\item Other: {[}Text field for specification{]}
\end{itemize}

\begin{center}\rule{0.5\linewidth}{0.5pt}\end{center}
\textbf{1. What is your gender?}
\begin{itemize}
	\tightlist
	\item Woman
	\item Man
	\item Non-binary
	\item Prefer not to disclose
	\item Other: {[}Text field for specification{]}
\end{itemize}

\textbf{2. What is your age?} {[}Open-ended response{]}

\textbf{3. Please specify the highest degree or level of education you have completed or are currently attending.}
\begin{itemize}
	\tightlist
	\item No high school degree
	\item High school graduate
	\item Diploma or the equivalent (for example, GED)
	\item Some college credit, no degree
	\item Professional degree
	\item Doctorate degree
	\item Prefer not to answer
	\item Other: {[}Text field for specification{]}
\end{itemize}

\textbf{4. Which of the following best describes your race/ethnicity?}
\begin{itemize}
	\tightlist
	\item White
	\item Black or African American
	\item American Indian or Alaska Native
	\item Hispanic or Latino
	\item Asian
	\item Native Hawaiian or Pacific Islander
	\item Prefer not to answer
	\item Other: {[}Text field for specification{]}
\end{itemize}

\textbf{5. How frequently do you give computer or technology advice (e.g., to friends, family, or colleagues)?}
\begin{itemize}
	\tightlist
	\item Never
	\item Rarely
	\item Sometimes
	\item Frequently
	\item Very frequently
	\item Prefer not to say
\end{itemize}

\textbf{6. Describe the community you live in.}
\begin{itemize}
	\tightlist
	\item Large city
	\item Suburb
	\item Small town
	\item Rural
	\item Prefer not to say
	\item Other: {[}Text field for specification{]}
\end{itemize}

\clearpage{}

\end{document}